\documentclass[aps,twocolumn,superscriptaddress]{revtex4}
\usepackage{amsfonts}
\usepackage{amsmath}
\usepackage{amssymb}
\usepackage{graphicx}
\usepackage{float}

\begin{document}

\title{Strong- versus Weak-Coupling Paradigms for Cuprate Superconductivity}
\author{Shahaf Asban}
\affiliation{Department of Physics, Technion-Israel Institute of Technology, Haifa 32000,
Israel}
\author{Meni Shay}
\affiliation{Department of Physics, Technion-Israel Institute of Technology, Haifa 32000,
Israel}
\affiliation{Department of Physics and Optical Engineering, Ort Braude College, P.O. Box
78, 21982 Karmiel, Israel}
\author{Muntaser Naamneh}
\affiliation{Department of Physics, Technion-Israel Institute of Technology, Haifa 32000,
Israel}
\author{Tal Kirzhner}
\affiliation{Department of Physics, Technion-Israel Institute of Technology, Haifa 32000,
Israel}
\author{Amit Keren}
\email{keren@physics.technion.ac.il}
\affiliation{Department of Physics, Technion-Israel Institute of Technology, Haifa 32000,
Israel}
\date{\today }

\begin{abstract}
Absolute resistivity measurements as a function of temperature from
optimally doped YBa$_{2}$Cu$_{3}$O$_{7-\delta }$, La$_{2-x}$Sr$_{x}$CuO$_{4}$%
, Bi$_{2}$Sr$_{2}$Ca$_{1}$Cu$_{2}$O$_{8-x}$, and (Ca$_{0.1}$La$_{0.9}$)(Ba$%
_{1.65}$La$_{0.35}$)Cu$_{3}$O$_{y}$ thin films are reported. Special
attention is given to the measurement geometrical factors and the
resistivity slope between $T_{c}$ and $T^{\ast }$. The results are compared
with a strong coupling theory for the resistivity derivative near $T_{c}$,
which is based on hard core bosons (HCB), and with several weak coupling
theories, which are BCS based. Surprisingly, our results agree with both
paradigms. The implications of these findings and the missing calculations
needed to distinguish between the two paradigms are discussed.
\end{abstract}

\maketitle

Two major discoveries were made at a very early stage in the study of
cuprate superconductivity. One was the Uemura relation for underdoped
samples \cite{UemuraPRL89}. This relation states that $T_{c}\propto \lambda
(0)^{-2}$, where $T_{c}$ is the superconducting transition temperature, and $%
\lambda (0)$ is the magnetic penetration depth at zero temperature. This
relation was found using the muon spin rotation ($\mu $SR) technique. The
second discovery was that for under doping and optimal doping, at
temperatures $T$ above $T^{\ast }$ the resistivity $\rho _{dc}(T)$ is a
linear function of $T$ \cite{AndoPRL04,BarisicCM}. Near optimal doping, $%
T^{\ast }$ is similar to $T_{c}$ and the linear relation extends down to $%
T_{c}$. Later on Homes extended the Uemura relation and showed that a
broader scaling holds for both underdoped, optimally doped, and overdoped
samples: $\rho _{s}(0)\propto \sigma (T_{c})T_{c}$ where $\rho _{s}(0)$ is
the superfluid density at zero temperature, and $\sigma (T_{c})=1/\rho
_{dc}(T_{c})$ is the conductivity at $T_{c}$ \cite{HomesPRB05}. This
observation was based on optical conductivity measurements. In many low
doping models, $\rho _{s}(0)\propto \lambda ^{-2}(0)$ \cite{BernardetPRB02}.
Therefore, the Homes law can be expressed as $\lambda ^{-2}(0)\propto \sigma
(T_{c})T_{c}.$ For both Homes and Uemura's laws to coexist, $\sigma (T_{c})$
must be universal for all underdoped cuprates.

Two kinds of theories address the Homes law. The first kind was provided by
Tallon \textit{et al.} \cite{TallonPRB06}, and the latter by Imry, Strongin,
and Homes \cite{ImryPRL12}, and by Kogan \cite{KoganCM13}. They predict%
\begin{equation}
\lambda (0)^{-2}=K\sigma (T_{c})T_{c}  \label{ISH}
\end{equation}%
where $K$ ranges from $120$, as in the original Homes law, to $K=240$. These
theories have a few elements in common. They assume weak coupling (WC), that
the resistivity arises mainly from disorder, that the BCS relation between
the superconducting gap and the critical temperature $\Delta \propto T_{c}$
is correct, and that the constant of proportionality (which varies a bit
between authors) is on the order of unity. The big advantage of these
theories is that they explain materials of all dopings. The disadvantage is
that they treat a compound such as optimally doped YBCO as a dirty
superconductor. In optimally doped YBCO, the resistivity extrapolates to
zero at $T\rightarrow 0$ (see below), which can only occur in the absence of
impurities. In fact, no experiment shows inhomogeneities in this compound
\cite{OferPRB06}. In addition, the weak coupling theories do not address the
temperature dependence of $\sigma (T)$ for $T>T_{c}$, which is very
different from simple metals \cite{GarlandPRL68}.

\begin{figure}[htbp]
\begin{center}
\includegraphics[trim=0cm 0cm 0cm 0cm,clip=true,width=9cm]{{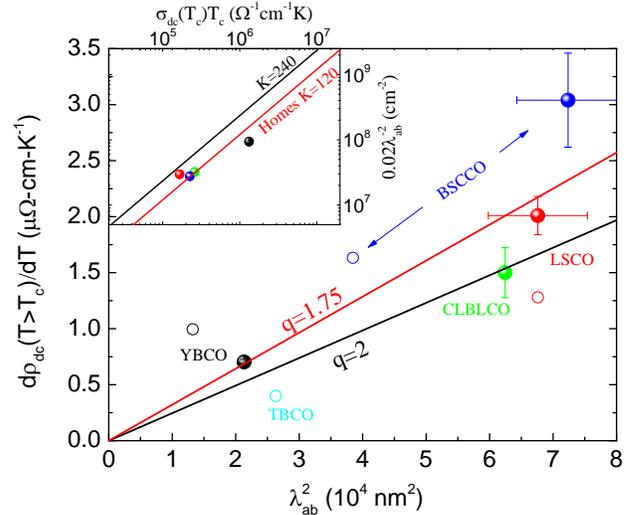}}
\end{center}
\caption{Solid symbols: The temperature derivative of the resistivity of
four different optimally doped cuprate films, at $T>T_{c}$, obtained by dc
measurements, as a function of their penetration depth. The solid lines show
the best linear fit to the data (that extrapolates to the origin) and the
prediction by the LA model. The values represented by open symbols are based
on single crystal measurements. The inset shows a Homes-type law on a
log-log scale generated from the same data. To get the same scales as Homes,
we are forced to multiply $\protect\lambda ^{-2}$ by 0.02. The original
Homes observation given by Eq.~\protect\ref{ISH} with $K=120$ and with the
largest $K=240$ obtained by WC predictions, are presented by solid lines.}
\label{main}
\end{figure}

The second theory was provided by Lindner and Auerbach (LA) \cite%
{LindnerPRB10}. They derived the relation $\rho _{dc}(T)=77.378\left( \frac{%
\lambda _{ab}\left( 0\right) }{q}\right) ^{2}\frac{K_{B}T}{\hbar c^{2}}$
using the hard core boson (HCB) model at half boson filling (optimal
doping); $q=2$ is the boson charge in units of $e$ and $K_{B}$ is the
Boltzmann constant. The HCB model is expected to be valid for temperatures
lower than $T^{\ast }$, where Cooper pairs are supposed to start forming in
the cuprates. This theory assumes a clean system and that the resistivity
arises from strong coupling (SC) between bosons. The LA derivation generates
the Homes law for optimal doping; it also captures the linear resistivity
and provides the coefficient of proportionality quantitatively. However, the
theory is not valid for an underdoped or overdoped compound, which is a
serious disadvantage. Due to impurities, the extrapolation to $T=0$ of $\rho
_{dc}(T)$ is finite in some cuprates. Therefore, it is more practical to
write the LA law in a differential form
\begin{equation}
\frac{d\rho _{dc}}{dT}(T>T_{c})=77.378\left( \frac{\lambda _{ab}\left(
0\right) }{q}\right) ^{2}\frac{K_{B}}{\hbar c^{2}}.  \label{MasterEq}
\end{equation}

\begin{figure}[H]
\begin{center}
\includegraphics[trim=0cm 0cm 0cm 0cm,clip=true,width=9cm]{{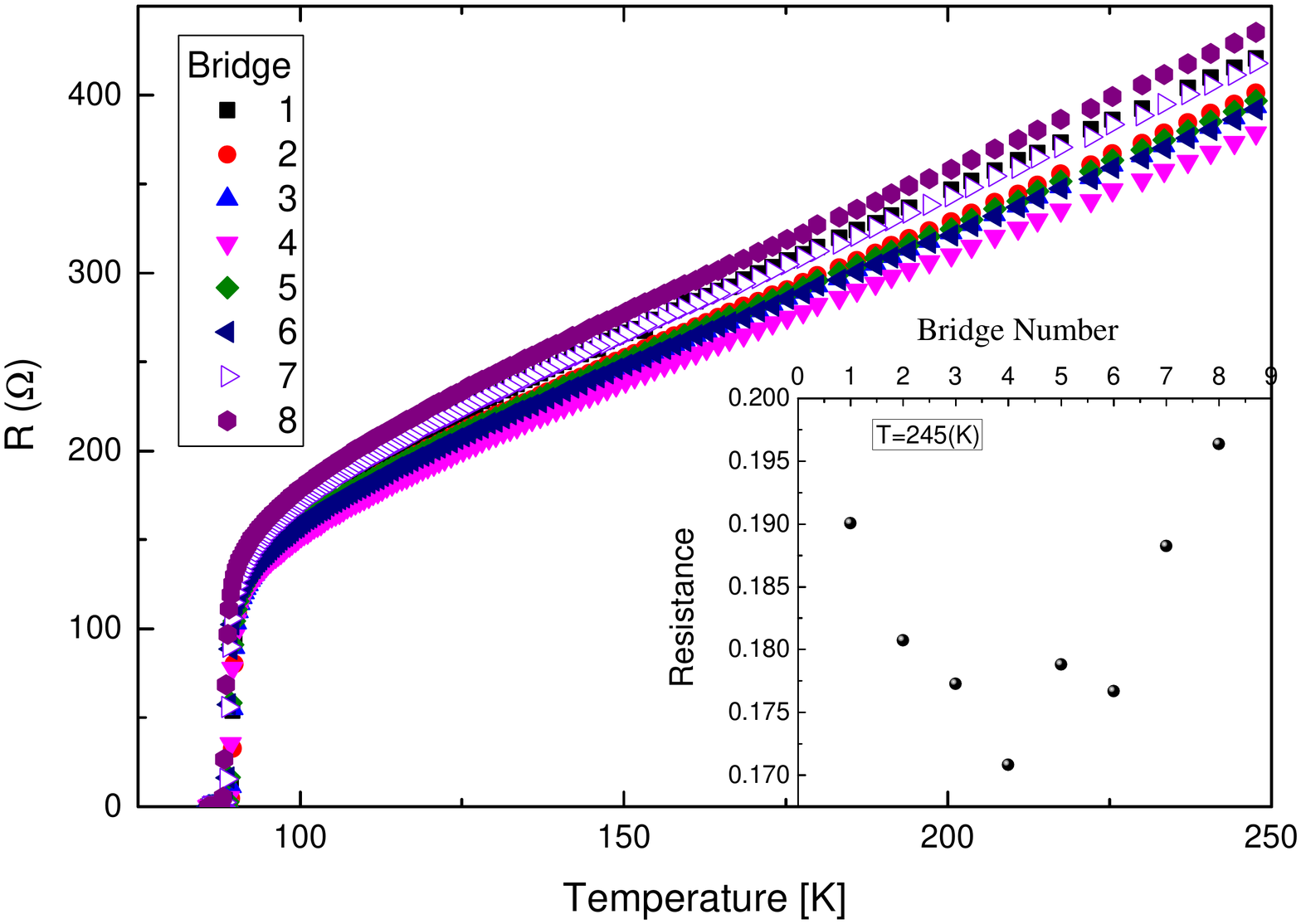}}
\end{center}
\caption{{}Measurements of resistance as a function of temperature in narrow
bridges of optimally doped YBCO. Bridges 1 and 8 are close to the edges of
the film. Bridges 3-6 are in the center of the film.}
\label{Bridge}
\end{figure}

In this work, we check both the WC and SC theories, as accurately as
possible, in the small region where both are valid, namely, optimal doping.
We use direct current (dc) resistivity versus $T$ measurements in films of
YBa$_{2}$Cu$_{3}$O$_{7-\delta }$ (YBCO), (Ca$_{0.1}$La$_{0.9}$)(Ba$_{1.65}$La%
$_{0.35}$)Cu$_{3}$O$_{y}$ (CLBLCO), La$_{2-x}$Sr$_{x}$CuO$_{4}$ (LSCO), and
Bi$_{2}$Sr$_{2}$Ca$_{1}$Cu$_{2}$O$_{8-x}$ (BSCCO). We take the geometrical
factors of the film into account and check their influence experimentally.
This allows us to determine $\rho _{dc}(T)$ in absolute value, and to
demonstrate that our results are indeed film-geometry independent. We then
compare $\frac{d\rho _{dc}}{dT}(T_{c})$ to $\lambda _{ab}^{2}(0)$ and $%
\sigma (T_{c})T_{c}$ to $\lambda _{ab}^{-2}(0)$, as in the SC and WC
theories, respectively. $\lambda _{ab}$ is taken from Refs. \cite%
{JacksonPRL00}, \cite{KerenSSC03}, \cite{lsco}, and \cite%
{ProzorovApplPhysLett00} respectively; the scatter in $\lambda _{ab}$ values
as provided by different authors is incorporated in the error bars as
described below. Our main results, given in Fig.~\ref{main}, are represented
by the solid symbols. For comparison we also show $\frac{d\rho _{dc}}{dT}%
(T>T_{c})$ for single crystals of YBCO, LSCO, BSCCO and Tl$_{2}$Ba$_{2}$CuO$%
_{6+\delta }$ (TBCO) taken from Refs.~\cite{AndoPRL04}, \cite{AndoPRL04},
\cite{WatanabePRL97}, and \cite{CarringtonPRB95} respectively, versus $%
\lambda _{ab}^{2}$ for single crystal taken from Refs. \cite{KieflPRB10},
\cite{lsco}, \cite{TallonPRB06} and \cite{TallonPRB06} respectively (open
symbols).

\begin{figure}[H]
\begin{center}
\includegraphics[trim=6cm 0cm 0cm 0cm,clip=true,width=12cm]{{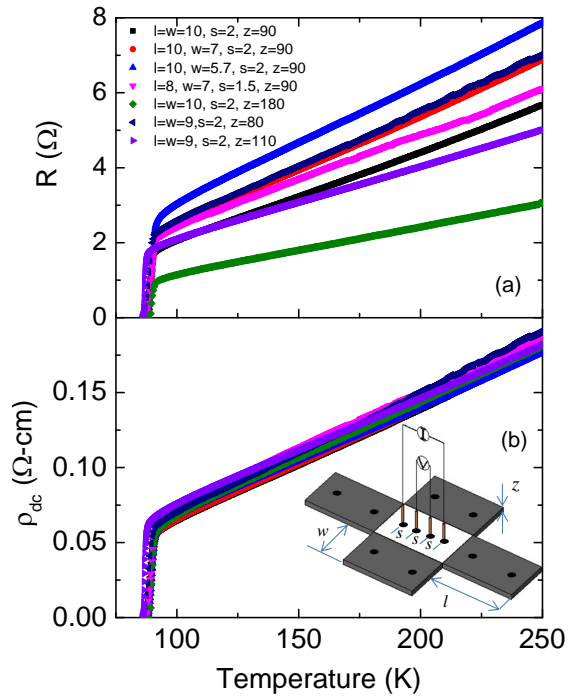}}
\end{center}
\caption{{}(a) Resistance vs temperature in films of different dimensions
and different distance between contacts. (b) The resistivity obtained by
using Eq.~\protect\ref{rhodc}. The resistivity is geometry-independent. The
inset in (b) shows the experimental set up and the set of current images
used to generate the correction factor calculated in \protect\ref{CfacCalc}.}
\label{Rtorho}
\end{figure}

In the case of the LA law, we fit our data to a straight line given by Eq.~%
\ref{MasterEq} with $q$ as a fit parameter. We find $q=1.75(15)$. The fit is
shown in Fig.~\ref{main}. We also depict in the figure the LA prediction
with $q=2$. The experimentally determined boson charge of $1.75(15)e$ is
very similar to theoretical charge of $2e$. It means that the HCB model is
self-consistent for the cuprates, and a very good starting point for
understanding the conductivity of optimally doped samples. We also present
our results as a Homes-type plot in the inset of Fig.~\ref{main}. Since
optical conductivity measures the plasma frequency which is proportional to $%
\lambda ^{-2}(0)$ it leaves one free parameter. To achieve the same scales
as Homes we multiply $\lambda _{ab}^{-2}(0)$ by $0.02$. This $2\%$
correction is due to the difference in the penetration depth and DC
conductivity as estimated by optical conductivity measurements and the
techniques used here \cite{DordevicNSR13}. With this scale we find that on a
log-log plot our data are not far from Homes', which are represented by the
solid line. We also show the WC theoretical predictions with $K=240$ in Eq. %
\ref{ISH}.

It seems that both WC and SC theories are in agreement with our experiment.
Another important piece of information is the indication of carriers with
charge $2e$ around the superconducting-insulator transition. This indication
comes from doping-temperature scaling relations of the resistivity \cite%
{BollingerNature11}. The emerging picture is that the superconducting state
in the cuprates is grainy, sometimes called Bose glass \cite%
{BollingerNature11}. The SC HCB model is a good starting point for
describing each grain. The normal metal between grains, in underdoped and
possibly overdoped samples, plays an important role in determining the
conductivity above the global $T_{c}$. This metal is best described by one
of the WC theories. However, at optimal doping one grain takes over the
entire sample. When this happens the conductivity is related only to the
superconducting properties, as Eq.~\ref{MasterEq} predicts.

\begin{figure}[htbp]
\begin{center}
\includegraphics[trim=1.5cm 0cm 0cm 0cm,clip=true,width=9cm]{{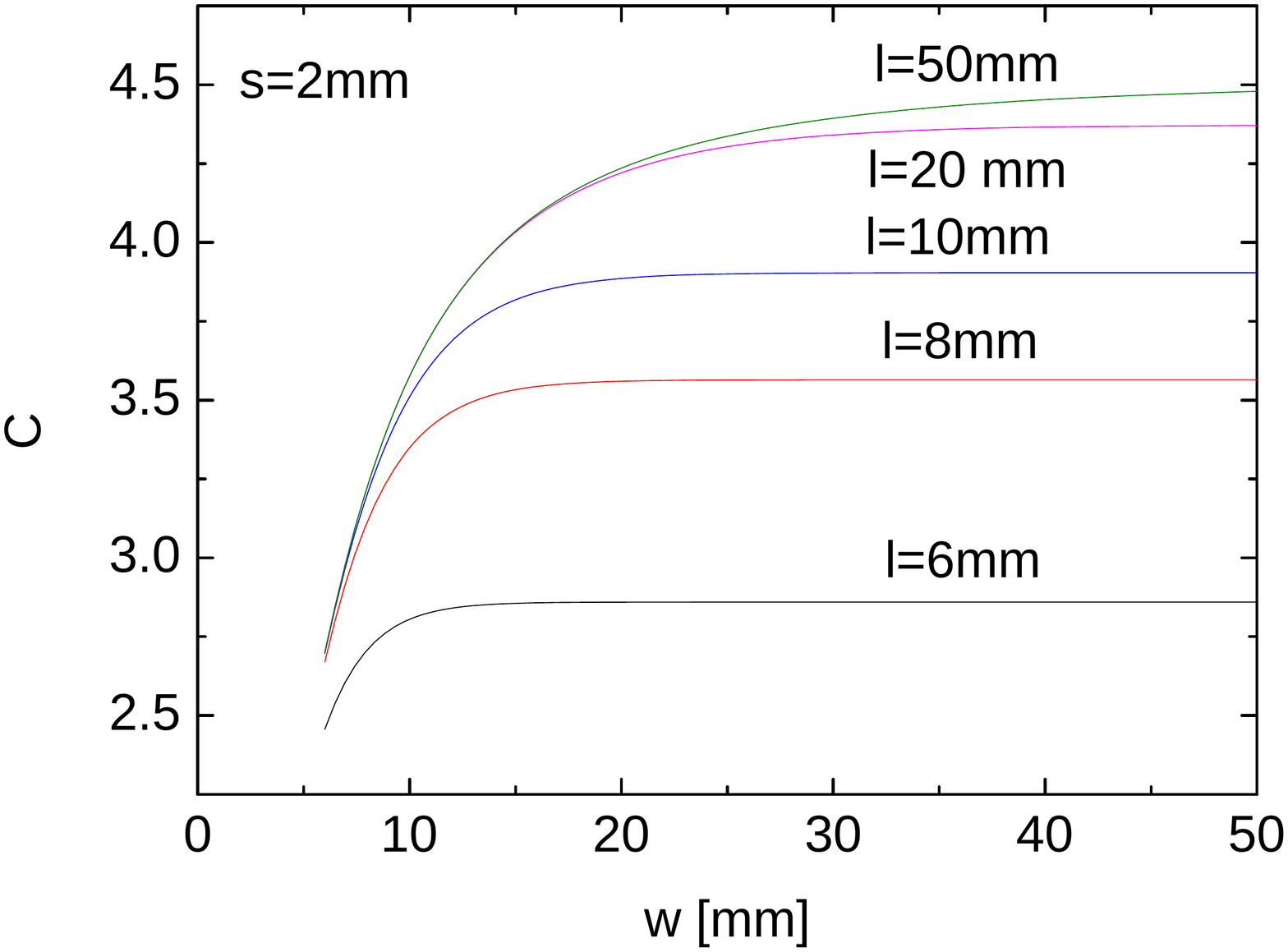}}
\end{center}
\caption{{}The geometrical factor, $C$, calculated in Eq.~\protect\ref%
{CfacCalc}, as a function of the width $w$ for various lengths $l$, and a
fixed distance between contacts $s$.}
\label{CorFacPlot}
\end{figure}

We now describe our experiment in more detail. A cardinal aspect of our
measurement is the determination of the absolute value of the resistivity
and resistivity derivative near $T_{c}$. One strategy is to use single
crystals, but in this case one does not know exactly which route the current
takes in the sample between contacts, and it is difficult to precisely
determine the resistivity. Therefore, such measurements are usually done by
preparing a film and patterning a bridge on it by ion milling. It is then
assumed that the resistance is dominated by the bridge. However, in high
temperature superconductors, close to $T_{c}$, the situation is not that
simple. Figure~\ref{Bridge} shows resistance measurement for a set of
identical bridges. For this and other measurements, we used films grown on a
$10\times 10$~mm$^{2}$ SrTiO$_{3}$ (STO) substrate with the $c\,$-axis
perpendicular to the film. Due to flux flow resistance, the transition
region from normal to the superconducting state is very rounded and it is
difficult to determine $\frac{d\rho _{dc}}{dT}(T\gtrsim T_{c})$. There is
also variation in the resistance between different bridges. This variation
is due to the film being less thick near the edges. The inset of Fig.~\ref%
{Bridge} shows the resistance at $T=245$~K as a function of bridge number.
Indeed, the first and last bridges are more resistive, but the middle ones
have very similar resistance. We therefore abandoned the bridge method, and
focused on wide film measurements which sample the film center and have very
sharp transitions, as shown in Fig.~\ref{Rtorho}. However, in this case
geometrical factors have to be taken into account when measuring resistivity
\cite{Smits}.

Our four-point probe measurement setup is shown in the inset of Fig.~\ref%
{Rtorho}(b). The two external contacts are used as the current source and
drain and the two internal contacts are the voltage probes. For a single
current source at the origin in contact with a two-dimensional (2D) infinite
conducting plane, the current density at a distance $r$ from the source is
given by $J=I/(2\pi r)$. The electric field on the conducting surface is set
by $J=\sigma E$. This leads to a logarithmic potential $V-V_{0}=-\frac{I}{%
2\pi }\rho _{dc}ln\,r$. In a current source ($a$) and drain ($b$), with
equal distance $s$ between all probes, the potential difference is $\Delta V=%
\frac{I}{\pi }\rho _{dc}ln\,2.$ For a finite sheet, the potential difference
is found by introducing an infinite number of images to the original current
source and drain, as shown by the spots on dark slabs in the inset of Fig.~%
\ref{Rtorho}(b) \cite{Smits}. This forces the current to run parallel to the
boundary. \noindent One then sums the potential from all images. The current
sources and drains, and their images, are located on a lattice given by $%
r_{nm}^{a}=(mw,s+nl)$ and $r_{nm}^{b}=(mw,nl-2s)$, where $w$ and $l$ are the
width and length of the film, respectively. \noindent \noindent The
potential difference between the two measured contacts is given by $\Delta
V=\rho _{dc}IC$, where
\begin{equation}
C=\frac{1}{2\pi }\underset{n,m}{\sum }\left( -1\right) ^{n}ln\left( \frac{%
\left( mw\right) ^{2}+\left( s+nl\right) ^{2}}{\left( mw\right) ^{2}+\left(
nl-2s\right) ^{2}}\right) .  \label{CfacCalc}
\end{equation}

\noindent $C$ as a function of $w$ for various $l$ and a typical $s$ is
shown in Fig.~\ref{CorFacPlot}. In our set up, $C$ is on the order of $3.5$.
Therefore, it is essential to check that Eq.~\ref{CfacCalc} is valid, as is
done below. Another important factor is the film thickness $z$, which is
measured by atomic force microscopy(AFM) as shown in the inset of Fig.~\ref%
{VI}. Each of the films waw measured from all sides, and, unless stated
otherwise, their thickness is $100\pm 5$~nm. The resistivity is given by
\begin{equation}
\rho _{dc}=\frac{1}{C}\frac{z\Delta V}{I}.  \label{rhodc}
\end{equation}%
Current simulations show that only $7\%$ of the total current passes close
to the edges of the films where the resistance is high by $7\%$ (see the
inset of Fig.~\ref{Bridge}). This leads to an error of less than $1\%$ on
the resistivity due to the thickness measurement.

To check the validity of Eqs.~\ref{CfacCalc} and \ref{rhodc}, we produced
YBCO films of various geometries and measured their resistance as presented
in Fig.~\ref{Rtorho}(a). The figure shows the resistance ($\Delta V/I$) of
seven different films with various heights $z$, widths $w$, lengths $l$ and
distances between contacts $s$, in units of millimeters. Figure~\ref{Rtorho}%
(b) depicts the resistivity $\rho _{ab}$ obtained by Eq.~\ref{rhodc}. The
resistivity is indeed geometry-independent and linear immediately above $%
T_{c}.$

\begin{figure}[H]
\begin{center}
\includegraphics[trim=4cm 0cm 0cm 0cm,clip=true,width=11cm]{{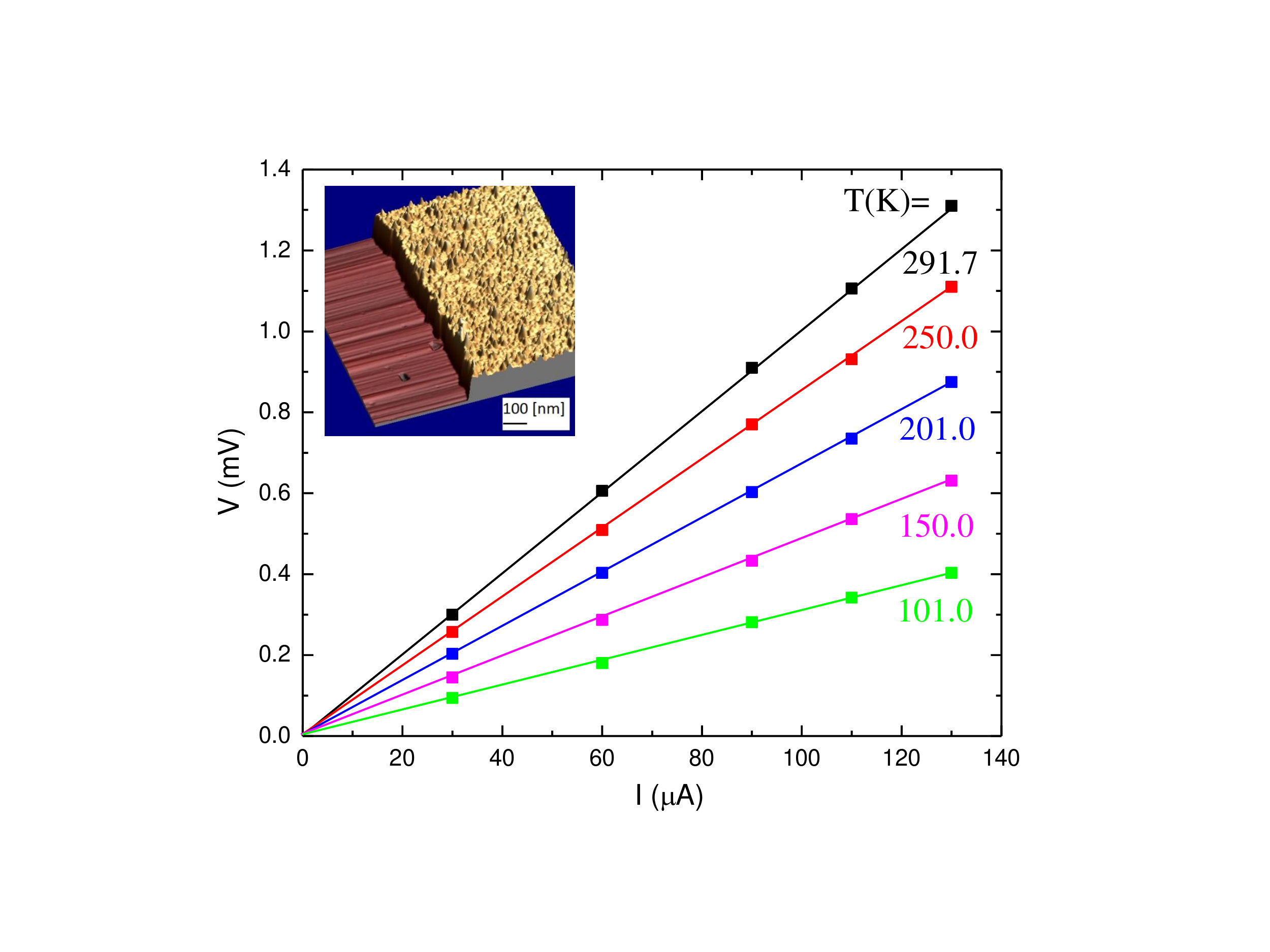}}
\end{center}
\caption{$V-I$ measurements of the YBCO film at different temperatures
demonstrating the ohmic behavior of the film. The inset shows an AFM image
of the film topography near an etched step. }
\label{VI}
\end{figure}

In Fig.~\ref{VI}, we show $V-I$ measurements of one of the YBCO films. In
the normal state, the films show ohmic behavior up to a current of 140~$\mu $%
A. Therefore, all our measurements are done in a current of 100$~\mu A$.

\begin{figure}[H]
\begin{center}
\includegraphics[trim=0cm 0cm 0cm 0cm,clip=true,width=9cm]{{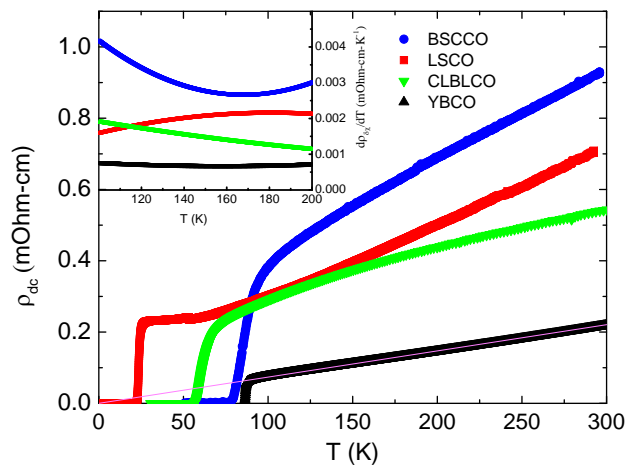}}
\end{center}
\caption{{}Resistivity versus temperature for four different optimally doped
cuprates. The inset shows the temperature derivative between 100 and 200~K,
which is above $T_{c}$ and below $T^{\ast }$ for all materials. The average
derivative is used in Fig.~\protect\ref{main}. The solid line demonstrates
that for YBCO $\protect\rho _{dc}(T\rightarrow 0)=0$.}
\label{AllMaterias}
\end{figure}

Finally, we present resistivity measurements in optimally doped films of
YBCO, LSCO, BSCCO, and CLBLCO in Fig.~\ref{AllMaterias}. A pure linear
behavior is observed only in YBCO, and, as expected, the resistivity
extrapolates to zero at zero temperature. In LSCO, the substrate reduces $%
T_{c}$ from the bulk value considerably, due to a mismatch in lattice
parameters. This lattice mismatch also reduce the $T_{c}$ of the other
compounds but not as much as in LSCO. To simplify our analysis, we focus on
the temperature range 100 to 200 K, which, for all materials, is higher than
$T_{c}$, higher than the region of fluctuating superconductivity, and lower
than $T^{\ast }$. In this temperature range, the reduction of $T_{c}$ in
LSCO is not relevant. In the inset of Fig.~\ref{AllMaterias}, we present the
first derivative of the resistivity as a function of temperature. As
expected, the derivative is a constant only for YBCO. For the other
materials, the derivative varies slowly with temperature. We treat the
derivative as a statistical variable and assign to each material an averaged
resistivity slope and standard deviation over the entire plotted range. The
standard deviation is used to generate the error bars. The summary of our
thermal derivative of the resistivity versus magnetic penetration depth
results is plotted in Fig.~\ref{main}. As mentioned before, the penetration
depth is taken from the literature. For optimally doped YBCO film, $\lambda
_{ab}=146\pm 3$~nm was determined in a theory-free method using slow muons
\cite{JacksonPRL00}. In this case, the value of $\lambda _{ab}$ agrees with
coated samples resonance (CSR) measurements, which is also a theory-free
method \cite{ProzorovApplPhysLett00}, and the error bar is known. For YBCO
crystal $\lambda _{ab}=115\pm 3$ was also measured with slow muons \cite%
{KieflPRB10}. For LSCO, there are only crystal measurements and all values
reported are scattered around $260\pm 15$ nm \cite{lsco}. For BSCCO, the $%
\lambda _{ab}=270\pm 15$~nm value was taken from CSR with its error bar \cite%
{ProzorovApplPhysLett00}. For BSCCO and TBCO crystals the value of $\lambda
_{ab}=196$ and $\lambda _{ab}=162$ respectively are from Ref.~\cite%
{TallonPRB06}. They have been measured by a few techniques but no error bar
is assigned. Finally, CLBLCO was measured only by standard $\mu $SR where
the determination of $\lambda _{ab}=250$~nm involves theoretical arguments
and the error bar is not known \cite{KerenSSC03}.

A comparison between our experimental results and both WC and SC theories
show that both are valid for optimally doped samples to some extent. To
distinguish between the two the WC theories should be extended to provide $%
\sigma (T>T_{c})$. Similarly, the SC theory should be broadened to include
the doping dependence of $\rho (T>T_{c}$). We believe that there is room for
a third theoretical approach that combines the two paradigms into one, in
order to account for the full doping and temperature variations. As for
optimal doping, the fact that the resistivity above $T_{c}$ is determined by
the superconducting quantity $\lambda _{ab}(0)$ only is an amazing property
of the cuprates.

We would like to thank Gad Koren for the use of his film preparation
laboratory. In addition we are grateful for helpful discussions with N. H.
Lindner, A. Auerbach, and Y. Imry. This research was funded by the Israeli
Science Foundation.

\end{document}